Topics: Misinformation, Echo Chambers, Bias, Polarization

# Right and left, partisanship predicts (asymmetric) vulnerability to misinformation

*Dimitar Nikolov, Alessandro Flammini, and Filippo Menczer*
Observatory on Social Media, Indiana University

We analyze the relationship between partisanship, echo chambers, and vulnerability to online misinformation by studying news sharing behavior on Twitter. While our results confirm prior findings that online misinformation sharing is strongly correlated with right-leaning partisanship, we also uncover a similar, though weaker trend among left-leaning users. Because of the correlation between a user's partisanship and their position within a partisan echo chamber, these types of influence are confounded. To disentangle their effects, we perform a regression analysis and find that vulnerability to misinformation is most strongly influenced by partisanship for both left- and right-leaning users.

## Research questions

- Is exposure to more partisan news associated with increased vulnerability to misinformation?
- Are conservatives more vulnerable to misinformation than liberals?
- Are users in a structural echo chamber (highly clustered community within the online information diffusion network) more likely to share misinformation?
- Are users in a content echo chamber (where social connections share similar news) more likely to share misinformation?

## Essay summary

- We investigated the relationship between political partisanship, echo chambers, and vulnerability to misinformation by analyzing news articles shared by over 15,000 Twitter accounts in June 2017.
- We quantified political partisanship based on the political valence of the news sources shared by each user.

- We quantified the extent to which a user is in an echo chamber by two different methods: (1) the similarity of the content they shared to that of their friends, and (2) the level of clustering of users in their follower network.
- We quantified the vulnerability to misinformation based on the fraction of links a user shares from sites known to produce low-quality content.
- Our findings suggest that political partisanship, echo chambers, and vulnerability to misinformation are correlated. The effects of echo chambers and political partisanship on vulnerability to misinformation are confounded, but a stronger link can be established between partisanship and misinformation.
- The findings suggest that social media platforms can combat the spread of misinformation by prioritizing more diverse, less polarized content.

## Implications

Two years since the call for a systematic "science of fake news" to study the vulnerabilities of individuals, institutions, and society to manipulation by malicious actors (Lazer et al., 2018), the response of the research community has been robust. However, the answers provided by the growing body of studies on (accidental) misinformation and (intentional) disinformation are not simple. They paint a picture in which a complex system of factors interact to give rise to the patterns of spread, exposure, and impact that we observe in the information ecosystem.

Indeed, our findings reveal a correlation between political partisanship, echo chambers, and vulnerability to misinformation. While this confounds the effects of echo chambers and political partisanship on vulnerability to misinformation, the present analysis shows that partisanship is more strongly associated with misinformation. This finding suggests that one approach for social media platforms to combat the spread of misinformation is by prioritizing more diverse, less polarized content. Methods similar to those used here can readily be applied to the identification of such content.

Prior research has highlighted an association between conservative political leaning and misinformation sharing (Grinberg et al., 2019) and exposure (Chen et al., 2020). The proliferation of conspiratorial narratives about COVID-19 (Kearney et al., 2020; Evanega et al., 2020) and voter fraud (Benkler et al., 2020) in the run-up to the 2020 U.S. election is consistent with this association. As social media platforms have moved to more aggressively moderate disinformation around the election, they have come under heavy accusations of political censorship. While we find that liberal partisans are also vulnerable to misinformation, this is not a symmetric relationship: consistent with previous studies, we also find that the association between partisanship and misinformation is stronger among conservative users. This implies that attempts by platforms to be politically "balanced" in their countermeasures are based on a false equivalence and therefore biased — at least at the current time. Government regulation could play a role in setting guidelines for the non-partisan moderation of political disinformation.

Further work is necessary to understand the nature of the relationship between partisanship and vulnerability to misinformation. Are the two mutually reinforcing, or does one stem from the other? The answer to this question can inform how social media platforms prioritize interventions such as fact-checking of articles, reining in extremist groups spreading misinformation, and changing their algorithms to provide exposure to more diverse content on news feeds.

The growing literature on misinformation identifies different classes of actors (information producers, consumers, and intermediaries) involved in its spread, with different goals, incentives, capabilities, and biases (Ruths, 2019). Not only are individuals and organizations hard to model, but even if we could explain individual actions, we would not be able to easily predict collective behaviors, such as the impact of a disinformation campaign, due to the large, complex, and dynamic networks of interactions enabled by social media.

Despite the difficulties in modeling the spread of misinformation, several key findings have emerged. Regarding exposure, when one considers news articles that have been fact-checked, false reports spread more virally than real news (Vosoughi at al., 2018). Despite this, a relatively small portion of voters was exposed to misinformation during the 2016 US presidential election (Grinberg et al., 2019). This conclusion was based on the assumption that all posts by friends are equally likely to be seen. However, since social media platforms rank content based on popularity and personalization (Nikolov et al., 2018), highly-engaging false news would get higher exposure. In fact, algorithmic bias may amplify exposure to low-quality content (Ciampaglia et al., 2018).

Other vulnerabilities to misinformation stem from cognitive biases such as lack of reasoning (Pennycook & Rand, 2019a) and preference for novel content (Vosoughi et al., 2018). Competition for our finite attention has also been shown to play a key role in content virality (Weng et al., 2012). Models suggest that even if social media users prefer to share high-quality content, the system-level correlation between quality and popularity is low when users are unable to process much of the information shared by their friends (Qiu et al., 2017).

Misinformation spread can also be the result of manipulation. Social bots (Ferrara et al., 2016; Varol et al., 2017) can be designed to target vulnerable communities (Shao, Hui et al., 2018; Yan et al., 2020) and exploit human and algorithmic biases that favor engaging content (Ciampaglia et al., 2018; Avram et al., 2020), leading to an amplification of exposure (Shao, Ciampaglia et al., 2018, Lou et al., 2019).

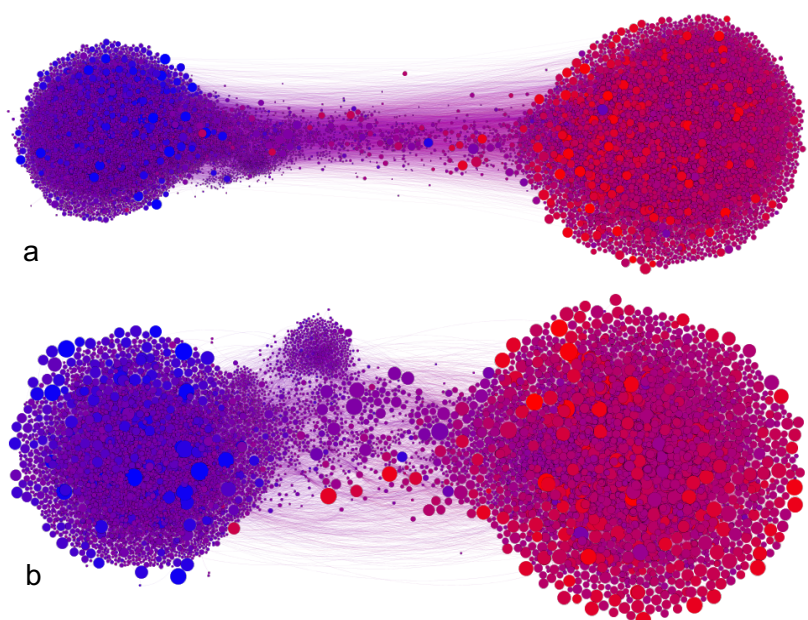

Figure 1. Echo-chamber structure of (a) friend/follower network and (b) diffusion network. Node color and size are based on partisanship and misinformation, respectively. See Methods for further details.

Finally, the polarized and segregated structure of political communication in online social networks (Conover et al., 2011) implies that information spreads efficiently within echo-chambers but not across them (Conover et al., 2012). Users are thus shielded by diverse perspectives, including fact-checking sources (Shao, Hui et al., 2018). Models suggest that homogeneity and polarization are important factors in the spread of misinformation (Del Vicario et al., 2016).

## Findings

Let us first reproduce findings about the echo-chamber structure of political friend/follower and diffusion networks on Twitter (Conover et al., 2011; 2012) using our dataset. Figure 1 visualizes both networks, showing clear segregation into two dense partisan clusters. Conservative users tend to follow, retweet, and quote other conservatives; the same is true for liberal users.

We want to test the hypothesis that partisanship, echo chambers, and the vulnerability to misinformation are related phenomena. We define a Twitter user's vulnerability to misinformation by the fraction of their shared links that are from a list of low-quality sources. Consistent with past research (Grinberg et al., 2019), we observe that vulnerability to misinformation is strongly correlated with partisanship in right-leaning users. However, unlike past work, we find a similar effect for left-leaning users (Figure 2a). Overall, we observe that right-leaning users are slightly more likely to be partisan and to be vulnerable to misinformation. The great majority of users on both sides of the political spectrum have misinformation scores below 0.5, indicating that those who share misinformation also share a lot of other types of content. In addition, we observe a moderate relationship between vulnerability to misinformation and two measures that capture the extent to which a user is in an echo chamber: the similarity among sources of links shared by the user and their friends (Figure 2b) and the clustering in the user's follower network (Figure 2c).

In Figure 3 we show that the three independent variables we analyze (partisanship, similarity, and clustering) are actually correlated with each other. User similarity and clustering are moderately correlated, as we might expect given that they both aim to capture the notion of embeddedness within an online echo chamber. Partisanship is moderately correlated to both echo chamber quantities.

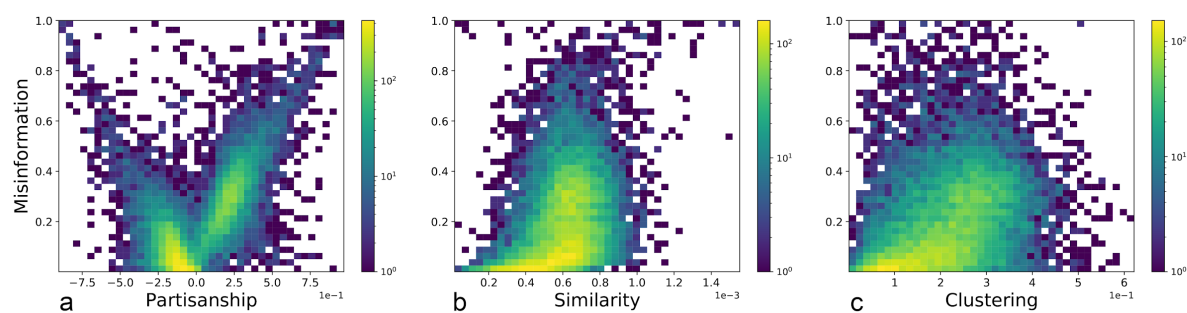

Figure 2. Relationship between misinformation and other variables. Color indicates the number of users. (a) Partisanship: the Pearson correlation is *r=0.64* for left-leaning users and *r=0.69* for right-leaning users. (b) Similarity: *r=0.33*. (c) Clustering: *r=0.30*. All correlations are significant (*p*<0.01).

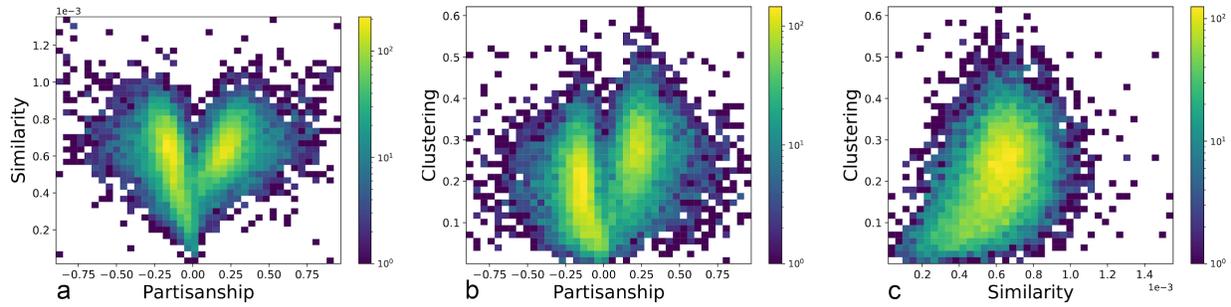

Figure 3. Relationship between independent variables. Color indicates the number of users. (a) Partisanship and similarity: Pearson correlation *r=0.45* for left-leaning users and *r=0.42* for right-leaning users. (b) Partisanship and clustering: *r=0.21* for left-leaning users and r=0.19 for right-leaning users. (c) Similarity and clustering: *r=0.40*. All correlations are significant (*p*<0.01).

Given these correlations, we wish to disentangle the effect of partisanship and echo chambers on vulnerability to misinformation. To this end, we conducted a regression analysis using vulnerability to misinformation as the dependent variable. The results are summarized in Table 1. We see that all independent variables except clustering for left-leaning users significantly affect the dependent variable (*p*<0.01). However, as shown by the coefficients and the adjusted increases in $R^2$, the effect is much stronger for partisanship.

| Variable | Left-leaning Users (Adjusted $R^2$ = 0.42) | | | Right-leaning Users (Adjusted $R^2$ = 0.48) | | |
|---|---|---|---|---|---|---|
| | Coefficient | *p* | Adj. $R^2$ increase | Coefficient | *p* | Adj. $R^2$ increase |
| Partisanship ($b_1$) | 0.435 | < 0.001 | 0.3 | 0.651 | < 0.001 | 0.4 |
| Similarity ($b_2$) | 0.047 | < 0.001 | 0.003 | 0.023 | < 0.01 | 0.001 |
| Clustering ($b_3$) | 0.047 | < 0.05 | 0 | -0.095 | < 0.001 | 0.002 |

Table 1. Regression results: effects of the three independent variables on the vulnerability to online misinformation for left- and right-leaning users. Along with the coefficients, we show *p*-values and adjusted $R^2$ increases.

Identifying the most vulnerable populations may help gauge the impact of misinformation on election outcomes. Grinberg et al. (2019) identified older individuals and conservatives as particularly likely to engage with fake news content, compared to both centrists and left-leaning users. These characteristics correlate with those of voters who decided the presidential election in 2016 (Pew Research Center, 2018), leaving open the possibility that misinformation may have affected the election results.

Our analysis of the correlation between misinformation sharing, political partisanship, and echo chambers paints a more nuanced picture. While right-leaning users are indeed the most likely to share misinformation, left-leaning users are also significantly more vulnerable than moderates. In comparing these findings, one must keep in mind that the populations are different --- voters in the study by Grinberg et al. versus Twitter users in the present analysis. Causal links between political bias, motivated reasoning, exposure, sharing, and voting are yet to be fully explored. To confirm that partisanship is associated with vulnerability to misinformation for both left- and right-leaning users, we conducted regression analysis comparing partisan (liberal or conservative) against moderate users. To this end, we grouped all users together and took the absolute value of their partisanship score as an independent variable. The results from this analysis (Table 2) show that partisan users are more vulnerable than moderate users, irrespective of their political leanings.

| Variable | All Users (Adjusted $R^2$ = 0.37) | | |
|---|---|---|---|
| | Coefficient | $p$ | Adj. $R^2$ increase |
| Partisanship ($b_1$) | 0.492 | < 0.001 | 0.214 |
| Similarity ($b_2$) | 0.280 | < 0.001 | 0.068 |
| Clustering ($b_3$) | 0.101 | < 0.001 | 0.002 |

Table 2. Regression results: effects of the three independent variables on the vulnerability to online misinformation for all users. Along with the coefficients, we show $p$-values and adjusted $R^2$ increases.

## Methods

### Dataset

We collected all tweets containing links (URLs) from a 10% random sample of public posts between June 1 and June 30, 2017 through the Twitter API. This period was characterized by intense political polarization and misinformation around the world and especially in the U.S. (Boxel at al., 2020). Major news events included the announcement of the U.S. withdrawal from the Paris Agreement; reinstatement of the immigration ban from Muslim majority countries; emerging details from the Muller investigation about Russian interference in the 2016 U.S. election; American and Russian interventions in Syria; James Comey's Senate Intelligence Committee testimony; the Congressional baseball shooting; U.S. restrictions on travel and business with Cuba; a special congressional election in Georgia; and continuing efforts to repeal Obamacare.

Since we are interested in studying the population of active online news consumers, we selected all accounts that shared at least ten links from a set of news sources with known political valence (Bakshy et al., 2015, see "Partisanship" below) during that period. To focus on users who are vulnerable to misinformation, we further selected those who shared at least one link from a source labeled as low-quality. This second condition excludes 5% of active right-leaning users and 30% of active left-leaning users. Finally, to ensure that we are analyzing human users and not social bots, we used the BotometerLite classifier (Yang et al., 2020) to remove likely bot accounts. This resulted in the removal of a little less than 1% of the accounts in the network. The user selection process resulted in the dataset analyzed in this paper, whose statistics are summarized in Table 3.

| Tweets | Retweets | Quote tweets | Accounts |
|--------|----------|--------------|----------|
| 1,398,552 | 847,934 | 14,108 | 15,056 |

Table 3. Dataset statistics

**Network Analysis**

We created two networks. First, we used the Twitter Friends API to build the follower network among the users in our dataset. Second, we considered the retweets and quotes in the dataset to construct a diffusion network. Both networks are very dense and, consistently with previous findings, have a deeply hierarchical core-periphery structure (Conover et al., 2012). Application of $k$-core decomposition (Alvarez-Hamelin et al., 2006) to the follower network reveals a core with $k$=940, meaning that each user has at least 940 friends and/or followers among other accounts within the network core. Since the network includes over 4 million edges, in Figure 1a we sample 10% (432,745) of these edges and then visualize the $k$=2 core of the largest connected component. The retweet/quote network has a $k$=21 core and includes 111,366 edges. In Figure 1b we visualize the $k$=2 core of the largest connected component.

**Partisanship**

To define *partisanship*, we track the sharing of links from web domains (e.g., *cnn.com*) associated with news sources of known political valence. For a source of political valence, we use a dataset compiled by Facebook (Bakshy et al., 2015), which consists of 500 news organizations. By examining the Facebook page of each news organization, the political valence score was computed based on the political self-identification of the users who liked the page. The valence ranges from −1, indicating a left-leaning audience, to +1, indicating a right-leaning audience.

We define the partisanship of each user $u$ as

$$P_u = \sum_s p(s|u)v_s$$

where $p(s|u)$ is the fraction of links from source $s$ that $u$ shares, derived from the Twitter data, and $v_s$ is the political valence of source $s$. In correlation and regression calculations, we take the absolute value $|P_u|$ for left-leaning users.

This definition of partisanship implicitly assumes that the sharing of a link implies approval of its content. While this is not necessarily true all of the time, we can reasonably assume it is true most of the time. To verify this, we sampled a subset of the accounts in our dataset (246 left- and 195 right-leaning users) and examined their profiles. Figure 4 visualizes the most common hashtags in the profile descriptions, confirming that our partisanship metric closely matches our intuitive understanding of left- and right-leaning users.

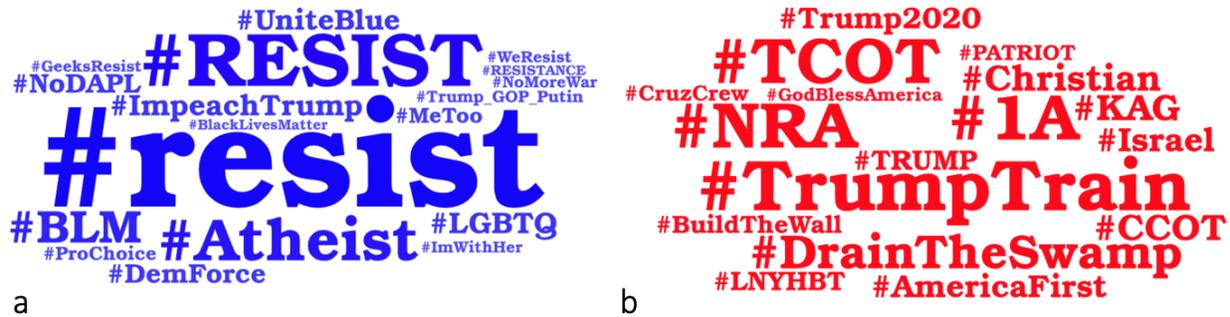

a          b

Figure 4. Most common hashtags appearing in the profile descriptions of (a) left-leaning users, and (b) right-leaning users. Font size is proportional to the number of occurrences across users.

**Misinformation**

To define *misinformation*, we focus on the sources of shared links to circumvent the challenge of assessing the accuracy of individual news articles (Lazer et al., 2018). Annotating content credibility at the domain (website) level rather than the link level is an established approach in the literature (Shao et al., 2018; Grinberg et al., 2019; Guess at al., 2019; Pennycook & Rand, 2019b; Bovet & Makse, 2019).

We considered a list of low-quality sources labeled by human judges (Zimdars et al., 2017). While the process employed for annotating sources is well-documented, the list is no longer maintained and its curation may be subject to the bias of the judges. Nevertheless, the list was current when our data was collected, and therefore has good coverage of low-credibility sources that were active at that time. More current lists of low-credibility domains do not allow for equal coverage because many sources have since become inactive. The annotation categories used to identify low-credibility sources in our dataset are shown in Table 4, along with statistics about their prevalence, as well as the numbers of links to news and other sources. Note that the total number of links shared is greater than the total number of tweets, for two reasons. First, a user can share multiple links in a single tweet; and second, most misinformation sources have multiple labels applied to them.

We extracted all links shared by each user, irrespective of whether they were from legitimate news, misinformation, or any other source. Each user $u$'s misinformation score $M_u$ is the fraction of all links they shared that were from sources labeled as misinformation.

**User Similarity**

To measure *similarity* between users, we construct the matrix of TF-IDF values for the user-domain matrix, with users analogous to documents (rows), and domains analogous to terms (columns). Thus, each user is represented by a vector of TF-IDF values indicating how strongly they are associated to each domain:

$$TFIDF(u, dom) = TF(u, dom) \times IDF(dom)$$

where $u$ is a user, $dom$ is a domain, $TF(u, dom)$ is the frequency with which $u$ shares from $dom$, and

$$IDF(dom) = log \frac{|U|}{|V(dom)|}$$

with $V(dom)$ being the subset of users who have shared from domain $dom$, out of the set of users $U$.

| Source | Tweets | Users |
|---|---|---|
| Extreme bias: Sources that come from a particular point of view and may rely on propaganda, decontextualized information, and opinions distorted as facts. | 55,064 | 9,833 |
| Clickbait: Sources that use exaggerated, misleading, or questionable headlines, social media descriptions, and/or images. These sources may also use sensational language to generate interest, clickthroughs, and shares. | 11,589 | 4,489 |
| Conspiracy: Sources that are well-known promoters of conspiracy theories. Ex: 9/11 conspiracies, chem-trails, lizard people in the sewer systems, birther rumors, flat earth 'theory,' fluoride as mind control, vaccines as mind control etc. | 24,964 | 4,801 |
| Fake news: Sources that entirely fabricate information, disseminate deceptive content, or grossly distort actual news reports. | 8,027 | 3,133 |
| Hate speech: Sources that actively promote racism, misogyny, homophobia, and other forms of discrimination. | 20,101 | 4,127 |
| Junk science: Sources that promote pseudoscience, metaphysics, naturalistic fallacies, and other scientifically dubious claims. | 427 | 237 |
| Rumor mill: Sources that traffic in rumors, gossip, innuendo, and unverified claims. | 3,885 | 1,364 |
| State propaganda: Sources in repressive states operating under government sanction. | 1,423 | 698 |
| Unreliable: Sources that may be reliable but whose contents require further verification or to be read in conjunction with other sources. | 12,546 | 5,244 |
| **Low-quality (Total)** | **138,026** | **15,056** |
| **News** | **463,409** | **15,056** |
| **Other** | **924,063** | **15,014** |

Table 4. Low-credibility, news, and other links shared in the dataset. Explanations of the low-quality labels are included (Zimdars et al., 2017), along with statistics about each link category. The 'other' category includes links to domains that are neither news nor low-quality sources.

To compute the similarity between two users, we take the cosine of the angle between their TF-IDF vectors. The similarity is defined in the unit interval. Finally, we take $S_u$ to be the average similarity between user $u$ and all of their friends. This measure quantifies how embedded a user is in their social network based on the content they and their friends share. A higher value indicates a homogeneous social network, which is one way to quantify an echo chamber.

**User Clustering**

To capture user *clustering* based on the follower network, we compute the clustering coefficient for each user as:

$$C_u = \frac{T(u)}{d_t(u)(d_t(u) - 1) - 2d_r(u)}$$

where $T(u)$ is the number of directed triangles through the user node $u$, $d_t(u)$ is the sum of in-degree and out-degree for $u$, and $d_r(u)$ is the reciprocal degree of $u$ (Fagiolo, 2007). This measure quantifies how densely interconnected a user's social network is. A dense follower network is another way to quantify an echo chamber.

**Regression Analysis**

To compare the different factors that may contribute to a user's misinformation score, we used multiple linear regression. Since the regression was performed for right- and left-leaning users separately, we took the absolute values of the partisanship scores. In addition, we took the z-score transform of each variable. The resulting regression equation is:

$$z_M = b_0 + b_1 z_{|P|} + b_2 z_S + b_3 z_C$$

**Acknowledgements**

We are grateful to Kai-Cheng Yang for his help in obtaining the bot scores used in this manuscript, and to Giovanni Luca Ciampaglia for constructive discussions.

**Funding**

This research is supported in part by the Knight Foundation and Craig Newmark Philanthropies. The sponsors had no role in study design, data collection and analysis, and article preparation and publication. The contents of the article are solely the work of the authors and do not reflect the views of the sponsors.

**Competing interests**
The authors declare no conflicts of interest.

**Ethics**
The Twitter data collection was reviewed and approved by the Indiana University IRB (exempt protocol 1102004860)

**Copyright**

**Data Availability**
Replication data for this study is available at https://doi.org/10.7910/DVN/6CZHH5 and https://github.com/osome-iu/misinfo-partisanship-hksmisinforeview-2021.